# Influence of system mass on the emission of intermediate mass fragments

Sukhjit Kaur, Supriya Goyal, and Rajeev K. Puri[*]
*Physics Department, Panjab University Chandigarh -160014, INDIA*
*email: rkpuri@pu.ac.in

## Introduction

The behaviour of hot and dense nuclear matter at the extreme conditions of temperature and density is a question of keen interest. It can be studied with the help of heavy-ion reactions at intermediate energies. At high excitation energies, the colliding nuclei may break into several small and intermediate size fragments followed by a large number of nucleons. This phenomenon is known as multifragmentation. Recently, Sisan et al. [1] studied the emission of intermediate mass fragments (IMFs) from central collisions of nearly symmetric systems using 4π-Array set up and found that the multiplicity of IMFs shows a rise and fall with increase in beam energy in the center-of-mass frame. They observed that peak $E_{c.m.}$ (energy at which the maximum production of IMFs occurs) increases linearly with system mass whereas a power law dependence proportional to $A^\tau$ has been observed for peak IMFs ($\tau \approx 0.7$). The rise and fall of IMFs with incident energy has also been established in several experimental studies [2,3]. Recently, Vermani et al. [4] carried out a theoretical study to check the mass dependence of emission of IMFs using soft equation of state (EoS) along with cugnon cross-section and MSTB method within the framework of QMD model. Their study was limited to symmetric nuclei only; we have extended the study to asymmetric nuclei also.

## The Model

The present study is carried out within the framework of Quantum Molecular Dynamics (QMD) model [4-9]. The QMD model is a n-body theory which simulates the heavy-ion reactions at intermediate energies on event by event basis. The fragments are identified using modified Minimum Spanning Tree method where in addition to spatial correlations fragments should also be bounded by binding energy:

$$\zeta_i = \frac{1}{N^f} \sum_{i=1}^{N^f} \left[ \frac{(\mathbf{p}_i - \mathbf{P}^{cm})^2}{2m_i} + \frac{1}{2} \sum_{j \neq i}^{N^f} V_{ij}(\mathbf{r}_i, \mathbf{r}_j) \right] < E_{bind} \quad (1)$$

We take $E_{bind}$ = -4.0 MeV if $N^f \geq 3$ and $E_{bind}$ = 0.0 otherwise. Here $N^f$ is the number of nucleons in a fragment and $\mathbf{P}^{cm}$ is center-of-mass momentum of the fragment. This is known as Minimum Spanning Tree with binding energy check (MSTB) method [4].

## Results and Discussion

We performed the simulations for the central reactions of $^{20}$Ne+$^{20}$Ne ($E_{lab}$ = 10-55 AMeV), $^{40}$Ar+$^{45}$Sc ($E_{lab}$ = 35-115 AMeV), $^{58}$Ni+$^{58}$Ni ($E_{lab}$ = 35-105 AMeV), $^{86}$Kr+$^{93}$Nb ($E_{lab}$ = 35-95 AMeV), $^{129}$Xe+$^{118}$Sn ($E_{lab}$ = 45-140 AMeV), $^{86}$Kr+$^{197}$Au ($E_{lab}$ = 35-400 AMeV) and $^{197}$Au+$^{197}$Au ($E_{lab}$ = 70-130 AMeV) using hard and soft equations of state along with different types of nucleon-nucleon (nn) cross-sections [8], momentum dependent interactions [9], and different widths of Gaussian.

In order to check the mass dependence of peak $E_{c.m.}$ (the energy where maximum fragments are emitted) and peak $<N_{IMF}>$, we plot, in fig. 1, the peak $E_{c.m.}$ and peak $<N_{IMF}>$ against the system mass. The peak $E_{c.m.}$ shows a linear dependence whereas a power law dependence of the form $cA^\tau_{tot}$ has been observed for peak $<N_{IMF}>$ [1,4], where $A_{tot}$ is the composite mass of the system. The power law parameter $\tau$ is found to be nearly equal to unity. We also noticed that QMD + MSTB is able to reproduce the experimental data [1,3] quite nicely. Further, different model ingredients such as width of Gaussian, different equations of state and cross-sections also affect the results. It is worth mentioning that IMF multiplicities in case of $^{20}Ne+^{20}Ne$ and $^{40}Ar+^{45}Sc$ are obtained by excluding the largest and second largest fragment respectively to get the accurate information about the system size dependence. The system size dependence is also observed for various fragments consisting of $A^{max}$, free-nucleons, light charged particles ($2\leq A\leq 4$), medium mass fragments ($5\leq A\leq 9$) as well as heavy mass fragments ($10\leq A\leq 44$). Interestingly, all these multiplicities of different fragments follow a linear power law except $A^{max}$.

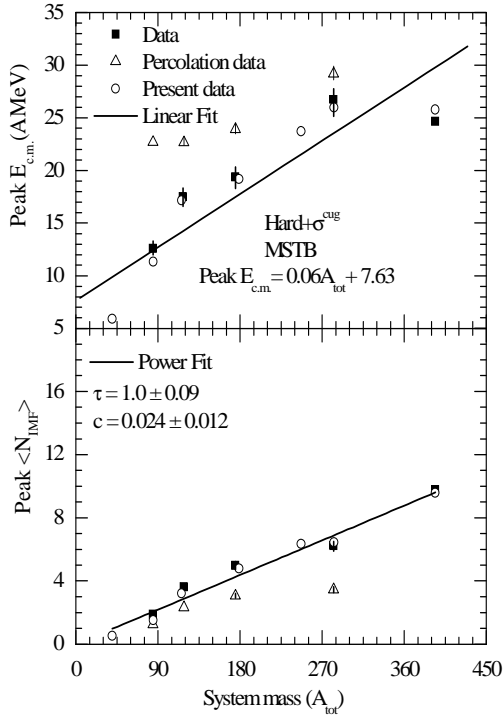

**Fig. 1** The peak $E_{c.m.}$ and peak $<N_{IMF}>$ versus composite mass of the system ($A_{tot}$). Comparison of model calculations (open circles) is made with experimental data (solid squares). The percolation calculations (open triangles) are also shown in fig. [1]. The experimental data for $^{197}Au+^{197}Au$ is taken from Ref [3].

## Summary

In the present study, we have simulated the central reactions of nearly symmetric and asymmetric systems over the entire periodic table for different values of energies using QMD model as event generator. These reactions are simulated for the different equations of state, nn cross-sections and different widths of Gaussians. We have observed that the multiplicity of IMFs shows a rise and fall with increase in beam energy in the center-of-mass frame as already predicted experimentally/theoretically. We have also studied the mass dependence of peak $E_{c.m.}$ and peak $<N_{IMF}>$. It has been observed that peak $E_{c.m.}$ increases linearly with system mass whereas a power law ($\alpha\ A^{\tau}_{tot}$) dependence has been observed for peak IMFs with $\tau \approx 1.0$. Our calculations using hard EoS along with cugnon cross-section and MSTB method are in good agreement with experimental data [1,3]. Similar power law dependence is also observed for various fragments.

## Acknowledgments

This work is supported by CSIR, Government of India.